\documentclass[aps,showpacs,showkeys,twocolumn]{revtex4}

\usepackage{amsmath, amsthm, amssymb,epsfig,alltt}

\def\a{\alpha}
\def\b{\beta}
\def\bt{\beta_T}

\def\p{\partial}

\def\t{\tau}

\def\s{\sigma}

\def\o{\omega}

\def\whatX{{\widehat{X}}}

\def\whatV{{\widehat{V}}}

\def\nn{\nonumber}

\def\2pap{2\pi\alpha^\prime}

\def\beq{\begin{eqnarray}}
 \def\eeq{\end{eqnarray}}
 \def\4pap{4\pi\a^\prime}

 \def\ap{{\a^\prime}}

\begin{document}


\title{Quantum Electron
Transport and Duality in \\
a One-Dimensional Interacting System}

\author{Taejin Lee \\~~\\
Department of Physics, Kangwon National University, \\
Chuncheon 200-701 Korea
\\
email: taejin@kangwon.ac.kr}




\maketitle

\centerline{\bf Astract}
We study the quantum electron transport in a one-dimensional interacting electron system, called 
Schmid model, reformulating the model in terms of the bosonic string theory on a disk. 
The particle-kink duality of the model is discussed in the absence of the external electric field and 
further extended to the model with a weak electric field. Using the linear response theory,
we evaluate the electric conductance both for both weak and strong periodic potentials
in the zero temperature limit. 
The electric conductance is invariant under the particle-kink duality.

\vskip 2cm

\section{Introduction}

In order to describe the dissipation quantum mechanically, 
Caldeira and Leggett \cite{caldeira83ann,caldeira83phy}
introduced a bath or environment, which consists of an infinite number of
harmonic oscillators, coupled to the system. The interaction with
the bath produces a non-local effective interaction. Later Schmid \cite{schmid,guinea} studied the
dissipative system in the presence of a periodic potential. The one dimensional dissipative
model with a periodic potential, called the Schmid model now, has been studied in conjunction 
with various interesting subjects such as Josephson junction arrays \cite{larkin,fazio,sodano}, 
the Kondo problem \cite{Affleck:1990by,Affleck:1990iv}, and 
the one-dimensional conductors \cite{kane:1992a}. 
Applications of the model includes also tunneling between Hall 
edge states \cite{kane2}, junctions of quantum wires \cite{Oshikawa:2005fh} 
and the decay of the unstable D-brane in string theory \cite{Sen:2002nu,senreview,Lee:2005ge,Hassel,Lee;2006,Lee;2007,Lee;2008}.

The Schmid model is found to exhibit non-trivial 
behavior; a phase transition, which is unlike one dimensional quantum mechanical systems with local
interactions only. Depending on the value of the friction constant, the phase diagram of
the system divides into two phases; the localized phase and the delocalized one.
In the delocalised phase, the periodic potential becomes an irrelevant operator and the electric 
conductance due to the particles can be easily calculated by the linear response theory.
At the critical point, the non-local effective action in one dimension is equivalent to 
a local action for a free fermion theory on a two dimensional disk and the periodic potential becomes a marginal mass operator on the boundary of the disk. The conductance of the model at the critical 
point is same as that of a free fermion theory. In the localized phase, the periodic potential 
becomes relevant and the particles are mostly localized at the minima of the periodic potential.
Then kinks, which are soliton solutions to the equation of motion, appear as new dynamical 
degrees of freedom. If the partition function is evaluated for the
general configuration of multi-kinks, it turns out to take the same form as that of the Schmid model
in the delocalized phase. This is the particle-kink duality, or the Schmid duality, which can be 
most efficiently represented by the string theory formulation of the model. 

The electric transport in the localized phase has been discussed in the literature ref.\cite{furusaki}, 
in the context of the spin-dependent Tomonaga-Luttinger model with a barrier potential, 
which is a generalized version of the Schmid model. However, the analysis was focused only 
on the electron transport
due to the tunneling from a potential minimum to an adjacent minimum. But the tunneling currents 
and the conductances are dependent upon the cutoff parameter and 
suppressed in the low energy (temperature) limit. We should note that the dynamical degrees of 
freedom in the localized are not electrons but the kinks or the array of multi-kinks. 

Reformulating the Schmid model in the framework of string theory \cite{callan90,callan91},
we evaluate the partition 
function in the presence of a spatially uniform, weak electric field. The particle-kink duality 
is extended to the model with a weak electric field and the electric conductance in the localized 
phase is calculated by using the linear response theory. The electron transport by the 
array of multi-kinks, represented by a dual field, may dominate over the transport due to the electron 
tunneling in the low energy limit.

\section{Particle-Kink Duality of the Schmid Model}

The dissipative system, describing quantum particles moving in one dimension in the 
presence of a periodic potential is called the Schmid model. Its quantum mechanical action
is given as follows
\beq 
S &=& \frac{\eta}{4\pi \hbar} \int^{\bt/2}_{-bt/2} dt dt^\prime 
\frac{\left(X(t) - X(t^\prime)\right)^2}{(t-t^\prime)^2} \nn \\
&& +  \frac{M}{2\hbar} \int^{\bt/2}_{-\bt/2} dt \dot X^2
- \frac{V_0}{\hbar} \int^{\bt/2}_{-\bt/2} dt \cos \frac{2\pi X}{a}. \label{action}
\eeq
The first term is responsible for the dissipation, and the second 
term is the usual kinetic term for a particle with mass $M$. 
Here $\eta$ is the frictional constant 
which measures the strength of the coupling the bath and the quantum
particle. The third term denotes the periodic potential. 
To produce the dissipative force in the classical limit, 
Caldeira and Leggett introduced a bath or environment which consists of an infinite 
number of harmonic oscillators which can be properly quantized. 
Assuming that the interaction between the 
bath and the system is linear and imposing the Ohmic condition for the spectral function 
of the oscillator frequencies, they found that the effective friction term could be
generated in the equation of motion. 
In the quantum theory if we integrate out the degrees of freedom of the bath, 
we obtain the non-local dissipative interation term in Eq.(\ref{action})
The period of time is chosen to be $\bt$, and $\bt \rightarrow \infty$ 
in the zero temperature limit. Since we are only 
interested in the long-time, low energy behavior of the system, we may ignore
the kinetic term, which only plays a role of regulator in the 
long-time analysis.

We can map the Schmid model to the string theory on a disk by identifying the time
as the boundary parameter $\s$ in string theory and scaling the field variable $X$ as follows:
$t = \frac{\bt}{2\pi} \s, \quad X \rightarrow \frac{a}{2\pi} X$.
Dropping the kinetic term, we have the string theory action for the Schmid model 
\beq \label{schmid}
S &=&  - \frac{1}{4\pi \ap} \int d\t d\s (\p_\t + i\p_\s) X (\p_\t - i\p_\s) X \nn\\
& & + \frac{V}{2} \int d\s\left(e^{iX} + e^{-iX}\right)
+ \frac{M}{2}\int d\s \left(\frac{dX}{d\s}\right)^2. 
\eeq
Here, the physical parameter of the two theories are identified as 
$\frac{\eta}{4\pi \hbar} \left(\frac{a}{2\pi}\right)^2 
= \frac{1}{8\pi^2 \ap}$.
When $\ap = 1$, this action conicides with the action of the rolling tachyon 
which depicts the decay of unstable D-branes in string theory. 

If $\ap >1$, the periodic potential is an irrelevant operator, which can be treated as 
a perturbation and dynamical degrees of freedom are particles. 
If $\ap<1$, the periodic potential term becomes a relevant operator. 
In the low energy regime the boundary 
potential dominates over the bulk string action and
the quantum particle is localized at the minina of the periodic potential. 
The physical degrees of freedom are kinks, which correspond to the tunneling of the quantum particles between adjacent minima of the periodic potential. (We also assume that $V/M \gg 1$.)
The kink is a solution to the 
equation of motion derived from the boundary action:
$X_K(\s) = 2 \, \arccos\left[-\tanh\left(\sqrt{\frac{V}{M}}\,\s\right)\right]$.
For the multi-kink solution we may take 
$X(\s) = \sum_{i=1}^n e_i X_K(\s-\s_i)$
where $e_i=1$ (kink) or $-1$ (anti-kink).

If the kinks are the principal physical degrees of freedom, we may rewrite the partition function as 
\beq
Z &=& \sum_n \sum_{\{e_i\}} \frac{1}{n!} \int d\s_1 \dots \int d\s_n y^n_0 \nn\\
&&\exp\left\{-\frac{\a}{2}
\sum_m \frac{1}{|m|} \left\vert \sum_{i=1}^n e_i e^{im\s_i} \right\vert^2\right\}
\eeq
where $\a = 1/\ap$ and 
$y_0 = \exp\left[-8\sqrt{MV}\right]$ is the instanton fugacity, or the 
tunneling matrix element between the adjacent minina. Here we make use of an apporximation
\beq
\int d\s e^{in\s} \frac{d\,}{d\s} X_K(\s-\s_i) = 2\pi e^{in\s_i}, \nn\\
\int d\s e^{in\s} X_K(\s-\s_i) = \frac{2\pi}{in} e^{in\s_i}, \quad n\not=0
\eeq
which is valid in the limit where $V/M \gg 1$. 
Finally, introducing a dual field, $\whatX$, 
\beq
\whatX (\s) = \sum_{m} \frac{1}{2\pi} \whatX_m e^{im \s}, 
\eeq
we may cast this partition function into the familiar form of the string theory 
\beq
Z &=& \sum_n \sum_{\{e_i\}} \frac{y^n_0}{n!} \int D[{\whatX}] \int \prod_{j=1}^n d\s_j \,\nn\\
&&\exp\left\{- \frac{1}{4\pi\a} \sum_{m} \frac{|m|}{2\pi} |\whatX_m|^2 + 
i \sum^n_{i=1} e_i \whatX(\s_i) 
\right\} \\
&=& \int D[\whatX] \exp\Biggl[- \frac{1}{4\pi\a} \int d\t d\s (\p_\t + \p_\s) 
\whatX (\p_\t -\p_\s) \whatX \nn\\
&&
+ \frac{\whatV}{2} \int d\s\left(e^{i\whatX} + e^{-i\whatX}\right)\Biggr] \nn
\eeq
where ${\whatV} = y_0 = e^{-8 \sqrt{MV}}$.
Thus, we find that if we rewrite the partition function 
in terms of a dual field in order to take the kinks as the physical degrees of freedom, 
the corresponding action takes the same form as the string theory action for the Schmid 
model Eq.(\ref{schmid}) but with some changes of parameters
\beq
\a \rightarrow \frac{1}{\a},~~~ V \rightarrow \exp\left[-8 \sqrt{MV} \right].
\eeq
This is called the Schmid duality. The region, $\a>1$ is mapped onto the region $\a<1$ and 
vice versa. The critical point, $\a=1$ is self-dual under the Schmid duality. The dual transformation
provides a framework for the perturbation theory defined in terms of the dual field in the region $\a>1$.

\section{Electron Transportation}

To study the electron transportation we consider a spatially uniform external electric field
$E(t) = - \partial  A/ \partial t$ with monochromatic frequency dependence $ E = i\omega A$.
The interation with the external electric field is described by the usual minimal coupling 
action
\beq 
S_E = \int^{\b_T/2}_{\b_T/2} J A dt =
\frac{e}{2\pi} \int^{\b_T/2}_{\b_T/2}   X \, E \,dt 
\eeq
where $J = \frac{e}{2\pi} \frac{d X}{dt}$ is the electric current. 
We calculate the electric current $\langle J \rangle$ induced by the interation with the external
electric field. The electric conductance is defined as
$\langle J \rangle = \s_{cond} E$ .
Note that since the Schmid model contains the non-local dissipative term, we cannot directly 
apply the conventional Kubo formula to evaluate the conductance. Here we employ the string theory 
formulation to evaluate the induced current and calculate the electric conductance using the 
linear response theory. \\

{\bf Electric Conductance in the Delocalized Phase}\\
We evaluate the conductance in the delocalized phase where $\a <1$. Since in this phase  
the periodic potential becomes an irrelevant operator, we may ignore it in the zero temperature limit. 
Thus, we may adopt the free string action on a disk to calculate $\langle J \rangle$
\beq 
\langle J \rangle &=& \lim_{\o \rightarrow 0+}  -i \frac{e^2}{\hbar} 
\frac{1}{4\pi^2} \int dt\, \langle
\dot{X}(0) X(t) \rangle_{E=0} \,e^{-i\o t} E  \nn\\
&=& \lim_{\o \rightarrow 0+} - \frac{e^2}{4\pi^2 \hbar} 
\frac{1}{\o} \int dt \langle \dot{X}(0) \dot{X}(t) \rangle_{E=0} e^{-i\o t} E . \label{current1}
\eeq
To calculate the electric conductance in the low energy limit, we take $\o \rightarrow 0+$.
The correlation function $\langle \p_t X(0) \p_t X(t)\rangle$ can be calculated by using the 
string theory formulation. Since $\s = 2\pi t/\b_T$, and $\langle \p_\s X(0) \p_\s X(\s) \rangle =
1/\left(2\a \sin^2\left(\frac{\s}{2}\right)\right)$,
we find 
\beq \label{current2}
\langle J \rangle =  \frac{e^2}{\a h} E, \quad \s_{cond} = \frac{1}{\a} \, \frac{e^2}{h}.
\eeq
The electric conductance in the delocalized phase obtained by the string theory formulation 
agrees with those of previous works \cite{furusaki,fisher;1985,kane:1992b}. 
The renormalization group flow of $V$ can be obtained by mapping the
Schmid model to the Thirring model \cite{Lee;thirring,Lee;2008}
\beq
V = V_0 \left[ \frac{\Lambda^2}{\mu^2}\right]^{\frac{\a-1}{2\a}}.
\eeq
Since we are only concerned the conductance in the low energy limit, we do not take into
account the contribution of the periodic potential to the mobility, which is negligible 
in the low enegy limit. \\

{\bf Electric Conductance in the Localized Phase}\\
In the off-critical region where $\a >1$, the periodic potential become strong and 
the particles are mostly localized at the minima of the periodic potential. 
The dynamical degrees of freedom of the system are no longer particles but kinks.
We must re-evaluate the partition function in the presence of the week electric field in this 
localized phase, taking into account the interaction term with the external electric field, 
$S_E$. The partition function reads as 
\beq
Z = \int D[X] e^{-S_0 - S_E}, \quad 
S_E = \frac{e}{2\pi} \int \frac{dX}{dt}\, A\, dt.
\eeq
For the multi-kinks configurations $X(\s) = \sum_{i=1}^n e_i X_K(\s-\s_i)$, the partition 
function is written as for the weak electric field 
\beq
Z &=& \sum_n \sum_{\{e_i\}} \frac{1}{n!} \int d\s_1 \dots \int d\s_n y^n_0 \nn\\
& & ~~\exp\Biggl\{-\frac{\a}{2}
\sum_m \frac{1}{|m|} \left\vert \sum_{i=1}^n e_i e^{im\s_i} \right\vert^2 \nn\\
&&~~~~~~~~~~~+ i \frac{e\bt}{4\pi^2}  \frac{E_p}{p} \sum_{i=1}^n e_i e^{-ip\s_i}
\Biggr\}
\eeq
Here we choose the external electric field with a monochromatic frequency dependence as 
$E(\s) = E_p e^{-ip\s}$ with a positive integer $p$. It is equivalent to choosing
\beq
E(t) = E_p e^{-i \o_p t}, \quad \o_p = 2\pi p / \bt \rightarrow 0+
\eeq 
in the zero temperature limit. Introducing the dual field $\whatX$ as before, 
$\whatX (\s) = \sum_{m} \frac{1}{2\pi} \whatX_m e^{im \s}$, 
we get the dual partition function in the presence of the presence of weak electric field 
\beq
Z &=& \sum_n \sum_{\{e_i\}} \frac{y^n_0}{n!} \int D[{\whatX}] \int \prod_{j=1}^n d\s_j \,\, \nn\\
& & 
\exp\Biggl\{- \frac{1}{4\pi\a} \sum_{m} \frac{|m|}{2\pi} |\whatX_m|^2 + 
\frac{e}{2\pi \a} \whatX_p E_p \nn\\
&&~~~~~~~~~~~~~~~~~~~~~~~+ i \sum^n_{i=1} e_i \whatX(\s_i) \Biggr\}.
\eeq 
It can be recast into the partition function of a string theory with the periodic boundary 
interaction
\beq
Z &=& \int D[\whatX] \exp\Biggl[- \frac{1}{4\pi\a} \int d\t d\s (\p_\t + \p_\s) 
\whatX (\p_\t -\p_\s) \whatX \nn\\
& & ~~ + \frac{e}{2\pi \a} \int d\s \whatX \, E 
+ \frac{\whatV}{2} \int d\s\left(e^{i\whatX} + e^{-i\whatX}\right)\Biggr].
\eeq
In the string theory formulation of the partition function, we see that the coupling of the 
dual field, which collectively represents 
the dynamical degrees of freedom in the localized phase, to the external 
weak electric field is $e/\a$. By examining the partition function in the presence of a weak electric 
field, we find that the particle-kink duality is extended as 
\beq
\a \rightarrow \frac{1}{\a},~~ e \rightarrow \frac{e}{\a}, ~~ V \rightarrow \exp\left[-8 \sqrt{MV} \right].
\eeq
Thus, the strong potential region $\a>1$ with with an electric charge $e$ is mapped onto the weak potential 
region $\a <1$ with an electric charge $e/\a$. Since we have the partition function for the dual field 
$\whatX$, the calculation of the electric conductance in the dual theory is straightforward: Following the same steps of the linear response theory
for the evaluation of the electric conductance in the delocalized phase 
Eqs.(\ref{current1},\ref{current2}), $J_{Dual} = \frac{e}{2\pi \a} \frac{d\whatX}{dt}$, 
\beq
\langle J_{Dual} \rangle &=& \left(\frac{e}{\a}\right)^2 \frac{1}{\frac{1}{\a} h} E = \frac{e^2}{\a h} E.
\eeq
Thus, the conductance in the zero temperature limit is formally invariant under the dual transformation.

\section{Conclusions}

The Schmid model, one dimensional interacting electron system with a periodic potential, 
has been studied for decades in connection with various phenomena in condensed matter physics. 
However, it has been discussed mostly in 
the framework of the perturbative Coulomb gas expansion. Here we discuss the model 
as a string theory on a disk with a boundary periodic potential. 
The particle-kink duality is more succintly presented and 
the electric conductance in the low temperature limit is conveniently evaluated. 
The particle-kink duality has been studied only in the absence of the external electric field
in the previous works. However, in order to examine the electric properties of the model in the 
localized phase where the periodic potential becomes strong, it is necessary to extend the particle-kink 
duality in the presence of the external electric field. The string formulation of the Schmid model
proves useful. We extend the duality in this paper by evaluating the partition function in 
the presence of the weak electric field. As a result we find that the coupling
of the dual field $\whatX$ to the electric field is a fraction of particle charge
and the electric conductance is formally 
invariant under the duality. In the localized phase the particles are localized near the 
minima of the periodic potential. But a new dynamical degree of freedom emerges, which 
is depicted by the dual field $\whatX$. By using the linear response theory and the sring 
theory formulation, we show that the dual field carries a fractional charge $\frac{e}{\a}$, $\a >1$ and 
it conducts the electric current. In the delocalized phase the conductance is larger and less  
in the localized phase than that of a free fermion, $\frac{e^2}{h}$.
Immediate extensions of this work may be found in various subjects in condensed matter physics and 
string theory, which include the spin-dependent Tomonaga-Luttinger model with a scattering potential 
\cite{furusaki} and the quantum dissipative Hofstadter system \cite{Lee;2009}.

\vskip 1cm

\noindent{\bf Acknowledgments:}
This study is supported by Kangwon National University.


\end{document}